# Asymmetry in in-degree and out-degree distributions of large-scale industrial networks

Jianxi Luo[1] and Daniel E. Whitney[2]

[1]Engineering Product Development Pillar, Singapore University of Technology and Design, 487372, Singapore
[2]Engineering Systems Division, Massachusetts Institute of Technology, Cambridge, MA 02139, United States

Many natural, physical and social networks commonly exhibit power-law degree distributions. In this paper, we discover previously unreported asymmetrical patterns in the degree distributions of incoming and outgoing links in the investigation of large-scale industrial networks, and provide interpretations. In industrial networks, nodes are firms and links are directed supplier-customer relationships. While both in- and out-degree distributions have "power law" regimes, out-degree distribution decays faster than in-degree distribution and crosses it at a consistent nodal degree. It implies that, as link degree increases, the constraints to the capacity for designing, producing and transmitting artifacts out to others grow faster than and surpasses those for acquiring, absorbing and synthesizing artifacts provided from others. We further discover that this asymmetry in decaying rates of in-degree and out-degree distributions is smaller in networks that process and transmit more decomposable artifacts, e.g. informational artifacts in contrast with physical artifacts. This asymmetry in in-degree and out-degree distributions is likely to hold for other directed networks, but to different degrees, depending on the decomposability of the processed and transmitted artifacts.

## I. INTRODUCTION

Many technologies and products today are not designed and produced by single integrated firms, but large-scale industrial ecosystems spanning many specialized but complementary firms. Such industrial ecosystems can be represented as networks of firms (as nodes) connected by inter-firm transactional relationships (as links), i.e. industrial networks, and analyzed using graph theory and network analysis techniques [1-4]. Despite the increasing awareness of industrial ecosystems as complex networks [5-7], there are few statistical analyses of industrial networks in the literature [7-10]. Complex network analysis may illuminate hidden factors that affect the working of design and production processes, and discover new network mechanisms that may be shared by general types of networks.

In an industrial network, individual firms design and produce different components and parts, and also exchange and assemble them into larger and larger systems [5,6,8-10]. The inter-firm exchanges of components and parts via transactions align firms for a shared functional goal. For example, firms in an automobile industrial network design and produce different parts of an "automobile"—an artifact whose basic function is to move humans and goods. An electronics industrial network creates artifacts such as computers, mobile phones and televisions whose basic function is to process information. Such system functions and the physical properties of exchanged artifacts across firms may condition the topologies of the networks [5,6,8-11]. In turn, network topologies may also influence how well industrial networks fulfill their functional goals [1,2,4,5,7,8,10,11]. However, comparatively little is known about the topologies of industrial networks and the physical antecedents and functional significance of possible topologies.

Many real-world natural, physical and social networks exhibit common topologies, such as "small-world [12-15]" and "scale-free [16-18]", which in turn give the networks specific systemic functions. *Small-world topology* means that any pair of nodes in a rather large network are connected only by a relatively short path as the result of high local clustering of neighbor nodes [1,12,13]. Small-world topology gives the network functional advantages in information-spreading or signal-propagation speed, computational efficiency, and synchronizability [1,12], but also the undesirable rapid propagation of infectious diseases [14,15]. *Scale-free topology* means a highly skewed degree distribution that decays as a power law [16-18], i.e. $p(k) \sim k^{-r}$, where $k$ is the degree of a node, $p(k)$ is the fraction of nodes in the network that have degree $k$, and $r$ is the exponent. The "power law" implies a small number of nodes have many more connections than most of the other nodes. The "scale-free" topology makes the network robust against random failures of nodes, but vulnerable to the failure of highly connected nodes [2,17,18].

Based on the analysis of the large-scale industrial network in the Tokyo industrial district, Nakano and White [8] argued that neither small-world nor scale-free topology can characterize the topology of that industrial network. Instead, they found that industrial network is strictly hierarchical and acyclic, consistent with Harrison White's hierarchical description of industrial networks [5,6], and proposed that a hierarchical topology characterizes the structure of industrial networks. In contrast, other studies [9,10] have found that the industrial network of the electronics sector in Japan in the early 1990s was only partially hierarchical and about 40 percent of inter-firm relationships were cyclic to each other. In this paper, we also briefly report the hierarchy degrees of our industrial networks.

Herein, we discover several previously unreported asymmetrical patterns in the in-degree and out-degree distributions of a collection of industrial networks, and provide explanations. Our findings provide more nuanced understanding of the topology of industrial networks and its antecedents, which can be potentially generalized for broader networks.

## II. DATA

Our data are on large-scale industrial networks for the design and production of automobiles and electronics in Japan. In an industrial network, firms are nodes, and supplier-customer transactional relationships are direct links. A link is created from firm $i$ to firm $j$ if $j$ is a major customer of $i$. Here, inter-firm transactions are only for physical components of automobiles and electronics. Such information for nodes and links is extracted from a well-known series of publications, "*The Structure of Japanese Auto Part Industry*" and "*The Structure of Japanese Electronics Industry*", based on regular surveys by Dodwell Marketing Consultants. The publications provide directories of identifiable firms in different industries and list major customers and suppliers for each firm. The directories are known for their completeness. Therefore, our data may support an approximate population analysis (more than a sample analysis). We had access to data published in 1983, 1993, and 2001 for automobiles; only one book was published for electronics, in 1993.

## III. METHODOLOGY

In this paper, we primarily explore the degree distributions of incoming and outgoing links of production firms. We plot the degree distributions on a log-log plot of the cumulative in-degree and out-degree distributions of the automobile and electronics industrial networks, and investigate the shape patterns of the distribution curves. We will begin with investigating how much the nodal degree distribution on a log-log plot exhibits a straight line, which suggests a power law distribution of nodal degrees. The power law of cumulative probability distribution translates the power-law degree distribution $p(k) \sim k^{-\gamma}$ with exponent $\gamma$ into $P(k) \sim \sum_{k'=k}^{\infty} k'^{-\gamma} \sim k^{-(\gamma-1)}$ with exponent $\gamma - 1$. $P(k)$ is the probability that a node has more than $k$ incoming or outgoing links. In this paper, we investigate in detail the specific deviations of the distribution patterns from the pure power law pattern, and provide explanations.

Despite the focus of this paper on degree distributions, we also briefly report the "small-worldness" and hierarchy degrees, to provide readers with a basic understanding of the structures of the industrial networks. We follow Watts and Strogatz [12] to measure the *small-worldness* of our networks, according to average nodal clustering coefficient and average shortest path length in the networks. We follow Luo and Magee [19] to measure the hierarchy degree of a network as the percentage of links that are not in any cycle.

## IV. RESULTS

The industrial networks are visualized in Fig. 1. Some descriptive network statistics are reported in Table I. There are a few general observations about these networks. First, while the automobile industrial networks have many more nodes and links, and a higher connectivity (measured by average degree $<k>$) than the electronics network, all the networks are quite sparse with $<k>$ far lower than the maximum possible $N$-1. Second, despite that the characteristic path lengths of all networks were rather small (in the range of 2.5 to 3), their clustering coefficients are all small, indicating that nodes are connected by short paths but not highly clustered. This suggests the lack of small-world effects, consistent with Nakano and White's finding [8]. Third, the electronics networks have many inter-firm transaction cycles, whereas each of the automobile networks has only one or two cycles as also shown in a previous study [9,10]. Fig.1 highlights the firms in cycles using blue triangles. The electronics industrial network does not comply to a pure hierarchy or directed acyclic graph, which earlier studies suggested as a general structure property of production networks [5,6,8].

TABLE I. Descriptive network statistics. $N$, number of nodes; $L$, number of links; $<k>$, average degree; $C$, average nodal clustering coefficient; $l$, average shortest path length; $H$, hierarchy degree, i.e. ratio of links not in cycles.

| Network | Automobile | | | Electronics |
|---|---|---|---|---|
| | 1983 | 1993 | 2001 | 1993 |
| $N$ | 356 | 679 | 627 | 227 |
| $L$ | 1480 | 2437 | 2175 | 648 |
| $<k>$ | 4.157 | 3.589 | 3.469 | 2.855 |
| $C$ | 0.023 | 0.018 | 0.019 | 0.035 |
| $l$ | 2.544 | 2.862 | 2.806 | 3.031 |
| $H$ | 0.997 | 0.999 | 0.999 | 0.596 |

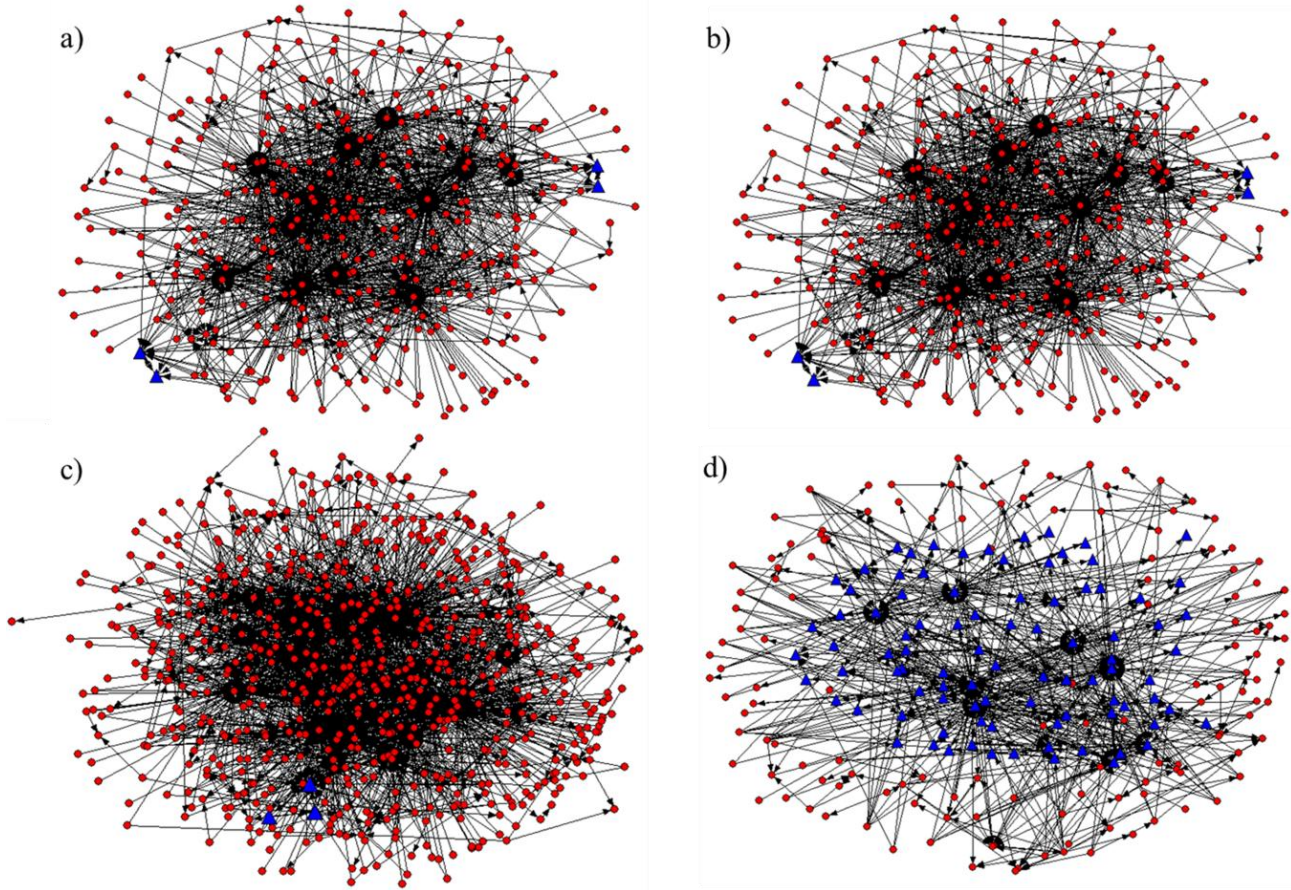

FIG. 1. Japanese industrial networks: a) automotive industrial network in 1983; b) automotive industrial network in 1993; c) automotive industrial network in 2001; d) electronics industrial network in 1993. Blue triangles represent nodes involved in inter-firm transaction cycles; red circles represent nodes that are not involved in any cycle.

Now we turn to the focus of this paper—a nuanced analysis of degree distributions of incoming and outgoing links of production firms. We found several common patterns as well as differences in the log-log plots of the cumulative in-degree and out-degree distributions of the automobile and electronics networks (Fig. 2).

The first common pattern is that all four networks show a power-law or scale-free regime, i.e., the straight-line regimes. A power-law out-degree distribution indicates several firms have many more customers than the remaining firms. Raw material suppliers, such as DuPont, have a high out-degree, as it supplies many types of firms, but a low in-degree as it is positioned extremely upstream of various value chains. Likewise, the power law of in-degree distribution indicates several firms have many more suppliers than other firms. This applies to system integrators such as Toyota, which has a high in-degree as it purchases materials, components, parts, and subsystems from many other types of firms, but a low out-degree as it is the most downstream in the production value chain. The functional implication of "scale-free" topology is that these industrial networks may be robust against random or accidental failures but are vulnerable to the failure of the most connected firms [17,18], such as Toyota generating various demands and DuPont supplying a wide range of other firms. The health of such key firms is crucial for the functioning of the entire industrial network [20,21].

The second common pattern is that out-degree distributions decay faster, showing a larger exponent or steeper slope of the regression lines than in-degree distributions. Particularly, out-degree distributions consistently cross in-degree distributions at $k^* \approx 10$, in our four networks. That is, when $k < 10$, it is easier for firms to add outgoing links (customers) than incoming links (suppliers). When $k > 10$, adding customers becomes more difficult than adding suppliers. The consistent crossing point $k^* \approx 10$ in all networks is particular and whether it is universal or a coincidence requires further research.

In explaining the growth of undirected networks, Amaral *et al* [22] and Mossa *et al* [23] suggested that the cost of connections of a node leads to cutoffs of the power-law regimes when $k$ is large. For directed networks, Braha and Bar-Yam [11]'s investigation of problem-solving networks revealed that in-degree distributions have sharp cutoffs that have substantially lower $k$ than those of the out-degree distributions, whereas their (in-degree and out-degree) scale-free regimes decay with similar exponents. They provide the explanation that the capacity for processing diverse incoming information is more constrained than that for disseminating the repeated information out to many receivers. Our observation of industrial networks is different in that 1) the scale-free regimes of out-degree distributions decay faster than in-degree distributions and that 2) out-degree distributions do not have obvious cutoffs. Despite these differences, the cost and capacity perspectives are also useful to explain our results.

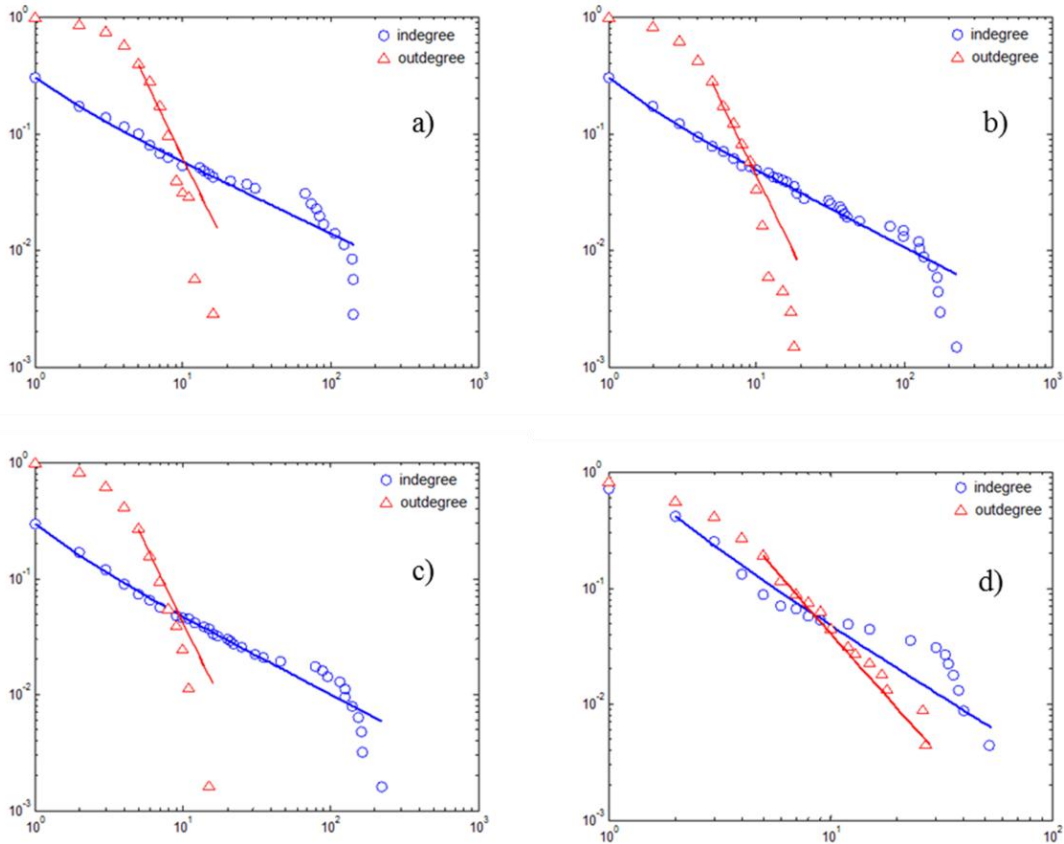

FIG. 2. Log-log plots of the cumulative in-degree and out-degree distributions of industrial networks: a) automotive industrial network in 1983; b) automotive industrial network in 1993; c) automotive industrial network in 2001; d) electronics industrial network in 1993. The horizontal axis is nodal in-degree or out-degree; the vertical axis is the cumulative probability of in- and out-degrees, $P_{in}(k)$ and $P_{out}(k)$, which are the probabilities that a firm has more than $k$ incoming and outgoing links.

In the industrial network context, the capacity for designing, producing and selling physical (instead of informational) artifacts to customers may be less constrained than that for acquiring, absorbing and synthesizing the artifacts from suppliers *only* when $k$ is small (< 10), but more constrained when $k$ is high (> 10). This may further imply the learning curve for designing, producing and transmitting products to others is steeper than that for absorbing and synthesizing the acquired products from other firms, thus reaching a limit faster.

The reverse asymmetry in in- and out-degree distributions between our observations and those of Braha and Bar-Yam [11] may be explained by the different nature of the processed and transmitted artifacts. It was information in Braha and Bar-Yam's networks, and physical components and parts in our networks. Processing and transmitting to a variety of receivers is easier for information than for physical artifacts. The major difference is that information artifacts are generally more modular [24], or "decomposable" as Herbert Simon put it [25], than physical artifacts.

The third common pattern across all four networks is that the scale-free regimes of in-degree distributions have cutoffs at $k > k^* \approx 10$, whereas out-degree distributions exhibit no cutoff. The cut offs of in-degree distributions can again be explained by the growing costs and limited

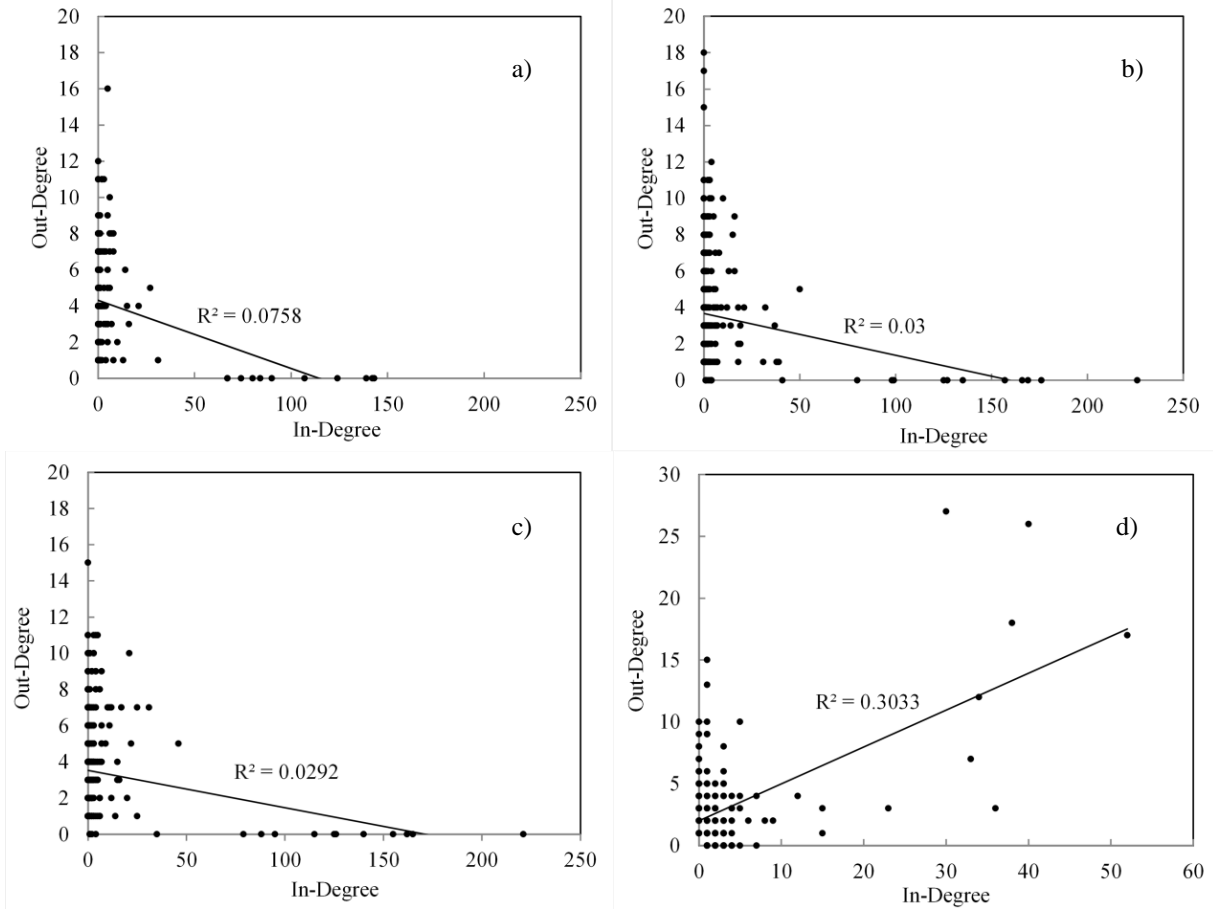

FIG.3. Correlation between in-degree and out-degree of firms in industrial networks: a) automotive industrial network in 1983; b) automotive industrial network in 1993; c) automotive industrial network in 2001; d) electronics industrial network in 1993. A node represents a firm, and its coordinates are in- and out-degrees.

capacity for adding new incoming links (suppliers) as $k$ increases.

Some topological differences also exist between automobile and electronics industrial networks. First, out-degree distributions decay faster in automobile networks than in the electronics network, whereas their in-degree distributions decay similarly. In other words, the decaying exponents of in- and out-degree distributions are more similar for the electronics network than automobile networks. Connecting this observation with the observed same exponents of the scale-free regimes of in-degree and out-degree distributions of information-processing networks [11], we speculate the difference in decaying exponents of in- and out-degree distributions may be related to the nature of the artifacts processed and exchanged in the networks. Detailed reasoning follows.

Functionally, an automobile moves people and goods, thus it processes a huge amount of energy and requires the use of many high-power technologies. High power processing creates difficult-to-anticipate side effects, such as noise, vibration and heat. To limit such secondary effects in integrating different components and parts, firms need to redesign and customize them for specific use of each customer [10,24], thus requiring significant efforts and costs to establish an additional outgoing link in the network. In general, automobile systems are highly integral; decomposability of the artifacts and processes is low; specificity and customization is crucial for designing and selling products.

Conversely, the general function of electronics is to process information. This function can be achieved using low power technologies, which in turn enable independence between the design and use of components. It is relatively easy to modularize and standardize electronic components and parts [10,24], and allows easy decomposition and integration of related design and production activities for different components and parts. The decomposability of electronics and related processes is relatively higher than that of automobiles. It is relatively easy to sell the modular and standardized electronics components with context-free specifications to a variety of customers (i.e. establishing many additional outgoing links). At the extreme, in information processing networks [11], the processed and exchanged artifact, i.e. information, is extremely decomposable.

This difference in physical properties of the artifacts processed and exchanged in different networks, e.g. high power vs. low power (and even purely informational), and decomposability vs. integrality, drives us to speculate that the difference in the required capacity for "designing, producing and selling" and "acquiring, absorbing and synthesizing" physical products is smaller, if the processed and exchanged artifacts involve lower physical power and are more modular or decomposable.

The same mechanism may also explain the second difference in the degree of correlation between the in-degree and out-degree of individual firms (Fig. 3). The out-degree and in-degree are relatively more correlated in the electronics network than in automotive networks, because the constraints faced by developing new outgoing and incoming links are more similar for firms in networks processing low-power and decomposable artifacts than those in networks processing higher physical power and less decomposable artifacts.

We notice that the automobile industrial networks have a set of special nodes with $k_{out}=0$ and $k_{in}>50$, which the electronics network does not have. They are the system assemblers, such as Toyota, Honda and Nissan in the very downstream of the production value chain, which have many suppliers but zero customers. We tested removing these special nodes from the automobile networks, and found in- and out-degree correlation coefficients remain extremely small. $R^2$ equals 0.0004, 0.0002 and 0.0131 for the automobile networks in 1983, 1993 and 2001 respectively. The conclusions above still hold.

Another difference lies in the cutoffs of in-degree distributions of automobile and electronics networks. The cutoff occurs at $k_{in}^+ \approx 10^{1.2}$ for the electronics network, lower than that for the automobile networks at $k_{in}^+ \approx 10^2$. This distinction may be related to the varied scales of these two industries and the products they produce. Automobiles are much larger-scale systems than electronics and contain many more components and parts to be outsourced and procured.

## V. CONCLUSIONS

This paper reveals several nuanced topologies in terms of the in-degree and out-degree distributions of industrial networks, which have implications to more general directed networks. This analysis is based on a sample of industrial networks, which are sparse and not highly clustered and some of which are only partially hierarchical. Our primary findings are on the asymmetries of the in-degree and out-degree distributions, consistent across these industrial networks. On that basis, we provide explanations to such asymmetries based on the physical natures of the artifacts being transacted among firms in the industrial networks.

We discover out-degree distribution decays faster than in-degree distribution and crosses it at a consistent degree $k^* \approx 10$ in all networks. We explain this asymmetry by the steeper learning curve for designing-producing-selling physical artifacts for others than that for acquiring-absorbing-synthesizing physical artifacts provided by others. We further observe that this difference in the decaying rates of in- and out-degree distributions is smaller when the processed and exchanged artifacts involve lower physical power and are more decomposable. The extreme is information-processing network in which such difference in in- and out-degree decaying rates was unseen [19] because information and information-processing process are highly decomposable.

The asymmetry of in- and out-degree distributions in all of our subject networks and the variation in the difference between in- and out-degree decaying rates across different types of networks in our sample are all discovered for the first time. Thus they are new to the general network analysis literature. It is also the first time that physics is used to explain the topology of complex economic transaction networks of firms. Therefore, our findings and analyses also contribute new understandings about the complex production ecosystems.

For future research, it will be interesting to explore and test if these topological patterns that we newly discovered and the functional and physical explanations that we provided hold for more general complex networks. In addition, one can explore alternative mathematical models than the power law to fit the empirically observed in- and out-degree distributions of industrial networks. Prior studies [26, 27] have suggested alternative distribution functions that fit well with certain empirical degree distributions with cutoffs.

[1] S. Strogatz, Exploring complex networks, Nature **410**, 268 (2001).
[2] M. E. J. Newman, The structure and function of complex networks, SIAM Rev. **45**, 268 (2003).
[3] R. Albert and A. L. Barabási, Statistical mechanics of complex networks. Rev. Mod. Phys. **74**, 47 (2002).
[4] M. E. J. Newman, A. L. Barabási, and D. J. Watts, *The Structure and Dynamics of Networks* (Princeton University Press, Princeton, NJ, 2006).
[5] H. C. White, Businesses mobilize production through markets: parametric modeling of path-dependent outcomes in oriented network flows, Complexity **8**, 87 (2002).

STRUCTURE AND DYNAMICS (2015)

[6] H. C. White, *Markets from Networks: Socioeconomic Models of Production* (Princeton University Press, Princeton, NJ, 2002).

[7] F. Schweitzer, G. Fagiolo, D. Sornette, F. Vega-Redondo, A. Vespignani, and D. R. White, Economic networks: the new challenges, Science **325**, 422 (2009).

[8] T. Nakano and D. R. White, Network structures in industrial pricing: the effect of emergent roles in Tokyo supplier-chain hierarchies, Structure and Dynamics **2**, 130 (2007).

[9] J. Luo, C. Y. Baldwin, D. E. Whitney, and C. L. Magee, The architecture of transaction networks: a comparative analysis of hierarchy in two sectors, Ind. & Corp. Change **21**, 1307 (2012).

[10] J. Luo, *Hierarchy in Industry Architecture: Transaction Strategy under Technological Constraints* (Doctoral Dissertation, Massachusetts Institute of Technology, Cambridge, MA, 2010).

[11] D. Braha and Y. Bar-Yam, Topology of large-scale engineering problem-solving networks, Phys. Rev. E **69**, 015113 (2004).

[12] D. J. Watts and S. H. Strogatz, Collective dynamics of 'small-world' networks, Nature **393**, 440 (1998).

[13] M. E. J. Newman, Models of the small world, J. Stats. Phys. **101**, 819 (2000).

[14] C. Moore and M. E. J. Newman, Epidemics and percolation in small-world networks, Phys. Rev. E **61**, 5678 (2000).

[15] M. E. J. Newman, I. Jensen, and R. M. Ziff, Percolation and epidemics in a two-dimensional small world, Phys. Rev. E **65**, 021904 (2002).

[16] P. D. J. de Solla, Networks of scientific papers, Science **149**, 510 (1965).

[17] A. L. Barabasi and R. Albert, Emergence of scaling in random networks, Science **286**, 509 (1999).

[18] A. Clauset, C. R. Shalizi, and M. E. J. Newman, Power law distributions in empirical data, SIAM Rev. **51**, 661 (2009).

[19] J. Luo and C. L. Magee, Detecting evolving patterns of self-organizing networks by flow hierarchy measurement, Complexity **16**, 53 (2011).

[20] J. Luo, Which industries to bail out first in economic recession? ranking U.S. industrial sectors by the Power-of-Pull, Econ. Sys. Res. **25**, 157 (2013).

[21] D. E. Whitney, J. Luo, and D. A. Heller, The benefits and constraints of temporary sourcing diversification in supply chain disruption and recovery, J. of Purch. & Supp. Manag. **20,** 238 (2014).

[22] L. A. N. Amaral, A. Scala, M. Barthélémy, and H. E. Stanley, Classes of small-world networks, Proc. Natl. Acad. Sci. USA **97**, 11149 (2000).

[23] S. Mossa, M. Barthélémy, H. E. Stanley, and L. A. N Amaral, Truncation of power law behavior in 'scale-free' network models due to information filtering, Phys. Rev. Lett. **88**, 138701 (2002).

[24] D. E. Whitney, Why mechanical design cannot be like VLSI design, Res. Eng. Des. **8**, 125 (1996).

[25] H. Simon, The architecture of complexity, Proc. Amer. Phil. Soc. **106**, 467 (1962).

[26] D. R. White, L. Tambayong, and N. Kejžar, Oscillatory dynamics of city-size distributions in world historical systems, in *Globalization as Evolutionary Process: Modeling, Simulating, and Forecasting Global Change*, edited by G. Modelski, T. Devezas, and W. R. Thompson (Routledge, London, 2008).

[27] D. R. White and L. Tambayong, City system vulnerability and resilience: oscillatory dynamics of urban hierarchies, in *Modelling and Simulation of Complex Social System*, edited by V. Dabbaghian and V. Mago (Springer, New York, 2014).